\documentclass[12pt]{article}
\usepackage{graphicx}

\newcommand\pubnumber{Article 1 in eConf C1304143}

\def\napoli{Center for Space Plasma and Aeronomic Research\\
University of Alabama in Huntsville, Huntsville AL 35805, U.S.A.}
\def\support{\footnote{on behalf of the GBM Team}}

\def\Title#1{\begin{center} {\Large #1 } \end{center}}
\def\Author#1{\begin{center}{ \sc #1} \end{center}}
\def\Address#1{\begin{center}{ \it #1} \end{center}}

\newcommand\pubblock{\rightline{\begin{tabular}{l} \pubnumber\\
          \end{tabular}}}
\newenvironment{Abstract}{\begin{quotation}  }{\end{quotation}}
\newenvironment{Presented}{\begin{quotation} \begin{center}
             PRESENTED AT\end{center}\bigskip
      \begin{center}\begin{large}}{\end{large}\end{center} \end{quotation}}

\def\beq{\begin{equation}}
\def\eeq#1{\label{#1}\end{equation}}
\def\eeqn{\end{equation}}

\def\beqa{\begin{eqnarray}}
\def\eeqa#1{\label{#1}\end{eqnarray}}
\def\eeqan{\end{eqnarray}}

\let\bar=\overbar

\def\Dslash{\not{\hbox{\kern-4pt $D$}}}
\def\dslash{\not{\hbox{\kern-2pt $\del$}}}

\def\msb{{\bar{\ssstyle M \kern -1pt S}}}

\begin{document}
\begin{titlepage}
\pubblock

\vfill
\Title{Variability Time Scales of Long and Short GRBs}
\vfill
\Author{ P. N. Bhat\support}
\Address{\napoli}
\Author{ }
\Address{}
\vfill
\begin{Abstract}
Gamma-ray bursts (GRB) are extremely energetic events and produce highly diverse light curves. Light curves are believed to be resulting from internal shocks reflecting the activities of the GRB central engine. Hence their temporal studies can potentially lead to an understanding of the GRB central engine and its evolution. The light curve variability time scale is an interesting parameter which most models attribute to a physical origin e.g., central engine activity, clumpy circum-burst medium, or relativistic turbulence. We develop a statistical method to estimate the GRB minimum variability time scale (MVT) for long and short GRBs detected by GBM. We find that the MVT of short bursts is distinctly shorter than that of long GRBs supporting the possibility of a more compact central engine of the former. We also find that MVT estimated by this method is consistent with the shortest rise time of the fitted pulses. Hence we use the fitted pulse rise times to study the evolution of burst variability time scale. Variability time coupled with the highest energy photon detected in turn related to the minimum bulk Lorentz factor of the relativistic shell emitted by the inner engine. 

\end{Abstract}
\vfill
\begin{Presented}
GRB 2013 \\
the Seventh Huntsville Gamma-Ray Burst Symposium \\
Nashville, Tennessee, 14--18 April 2013
\end{Presented}
\vfill
\end{titlepage}
\def\thefootnote{\fnsymbol{footnote}}
\setcounter{footnote}{0}

\section{Introduction}

Despite exciting progress in observations, the nature of gamma-ray bursts (GRBs) remains a big puzzle~\cite{bhat11}. The temporal structure of GRBs exhibits diverse morphologies.
Most bursts are highly variable with a variability time scale significantly smaller than the overall duration, $T_{90}$. It has already been shown that the $T_{90}$, reflects directly the length of time that the Òinner engineÓ operates and the observed temporal variability reflects variability in the Òinner engineÓ~\cite{sari97a}. Clumpy circum-burst medium~\cite{dermer99}, or relativistic turbulence~\cite{zhang11} are some of the other possible physical processes responsible for the observed variability in GRB prompt emission light curves. 

Variability in GRB light curves is observed on various time scales. In this paper we present a new statistical method of estimating the minimum value of such variability time scales of a GRB and relate it to the minimum value of the fitted pulse rise time. This, in turn, can be related to the minimum Lorentz factor of the relativistic shells emitted by the central engine. The evolution of the minimum Lorentz factor is then related to the spectral evolution of the GRBs.

\section{Observations}
Gamma-ray Burst Monitor (GBM) onboard the {\it Fermi} satellite detects and locates GRBs at the rate of about 2 GRBs in 3 days. The time tagged event data (TTE) produced with a temporal precision of 2 $\mu$s and a spectral resolution of 128 channels are used for this analysis. GRB light curves with an optimum bin-width are decomposed using the simplest asymmetric pulses defined by a lognormal function superposed over a smooth background fitted to a quadratic function ~\cite{bhat12}. The goodness of fit to the actual GRB light curve is tested using Pearson's chi-square parameter normalized to the number of degrees of freedom.  The pulse shape constants are then derived from the best fit parameters. The advantage of this algorithm is that it converges even in the presence of multiple overlapping pulses to deconvolve complex highly variable light curves.

\section{The Method}
\begin{figure}[htb]
\centering
\includegraphics[height=3.05in]{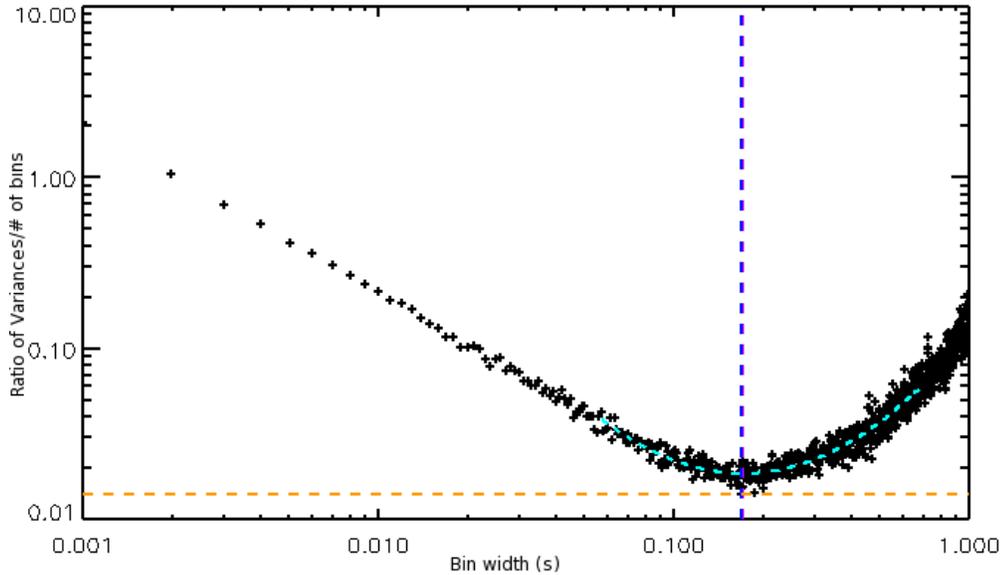}
\caption{: Variation of the ratio of the variances per bin to the histogram bin-width. At very fine bin-widths the GRB signal is indistinguishable from background fluctuations and hence the ratio decreases monotonically with increasing bin-width. At larger bin-widths the signal is clearly visible from the background and hence the ratio per bin starts increasing. The  bin-width at the turn over is defined as the minimum variability time scale where the bin-width is expected to be optimum. Cyan dashed line shows a fitted parabola around the minimum that has a minimum at a bin-width indicated by the vertical dashed line in blue.}

\label{fig:mvt}
\end{figure}

A new statistical method has been developed to estimate the MVTs of GRBs.  We identify the prompt emission duration of a GRB and an equal duration of background region selected from that part of the light curve before the trigger and after the GRB ends. Very high resolution GBM triggered GRB light curves are chosen for this purpose. We often start with a GRB light curve with a bin width of 100 $\mu$s. If the variability time is in 100s of ms then the chosen initial bin-width is 1 ms. We then derive a differential of each light curve and compute the ratio of the variances of the GRB and the background segments. This ratio divided by the bin-width. This process is repeated by incrementing the bin-width by the initial value until a maximum bin-width of 1 s is reached. The ratio of variances per bin-width is plotted as a function of bin-width in figure \ref{fig:mvt}.  The ratio in figure \ref{fig:mvt} initially falls as ${1} \over {bin-width}$ as seen in the figure. Simulations also show that for pure Poisson fluctuations this ratio falls monotonically with increasing bin-width until the real GRB signal becomes significant at a bin-width called the minimum variability time. 

The region around the minimum is fitted with a parabolic function (indicated by cyan dashed line) and the bin-width at the minimum is estimated. This bin-width where the minimum occurs is defined as the minimum variability time scale (indicated by the vertical blue dashed line). The error on the minimum variability time scale is the resulting error propagated from the parabolic fit parameters.

\begin{figure}[htb]
\centering
\includegraphics[height=3.05in]{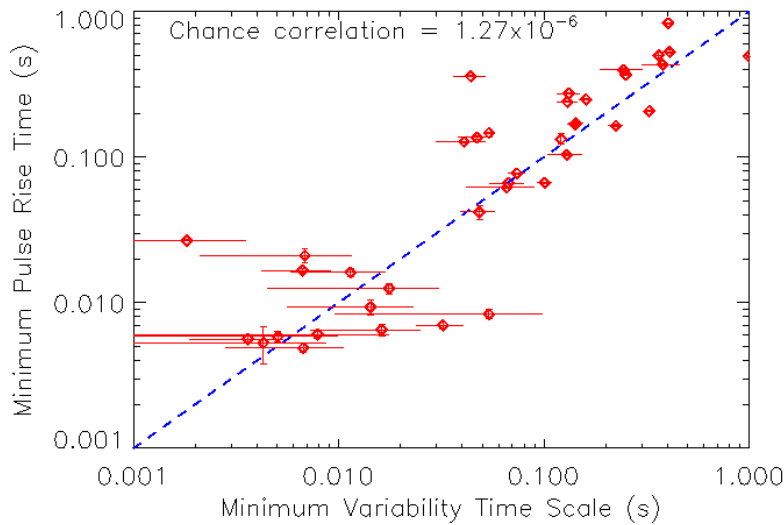}
\caption{: Shows a plot of the MVTs estimated by (a) the statistical method described
in figure 1 and by (b) using minimum of the rise time of the fitted pulses to the GRB
Prompt emission light curves. Over MVT values stretching 3 decades 
the two methods agree statistically. The dashed blue line is the line representing
Minimum Rise Time = MVT. It may be noted that at low MVT values the statistical errors are 
larger. The chosen GRB sample consists of both long and short GRBs.}

\label{fig:rt_mvt}
\end{figure}

Figure  \ref{fig:rt_mvt} shows a plot of minimum of the rise times of the fitted pulses as a function of the MVT of the same GRB. The linearity of the dependence  of minimum pulse rise time with MVT irrespective if the GRB duration demonstrates that the MVT does indeed represent the minimum variability of the GRB. In addition, the MVT estimated by an independent method by MacLachlan {\it et al.}~\cite{mach12} has been demonstrated to be linearly correlated with the minimum rise time of the fitted lognormal pulses to the GRB light curves.

\begin{figure}[htb]
\centering
\includegraphics[height=3.05in]{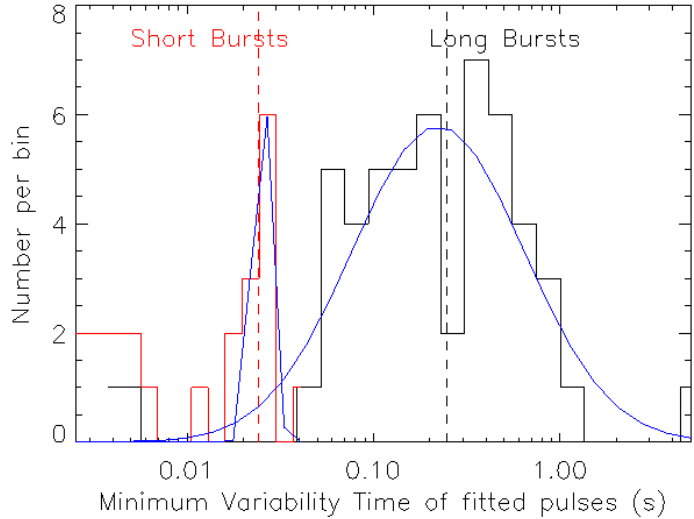}
\caption{: The minimum variability time scales are characteristic features of burst 
type. The MVTs of long (black histogram) and short (red histogram) form two 
separate distributions showing that the long and short GRBs have intrinsically 
different values of MVTs. The median values (shown as vertical dashed lines) are 
0.024 and 0.25  respectively for long and short GRBs. 
}

\label{fig:ls_mvt}
\end{figure}
\section{Results}

Figure \ref{fig:ls_mvt} shows a plot of minimum variability times for short (red histogram) and long (black histogram) GRBs. There seems to be a clear separation of variability times of short and long bursts with their respective median values differing by a factor of $\sim$10. Each of the histograms is fitted to independent lognormal functions (shown in blue). The median MVTs of long and short GRBs are 0.25\,s and 0.024\,s respectively (dashed vertical lines in figure \ref{fig:ls_mvt}). This is consistent with the previous conclusion that the central engines of shorts GRBs are perhaps more compact than those of long GRBs~\cite{bhat12,sari97b} . This is also consistent with the observed correlation between variability times and GRB durations by Gao {\it et al.}~\cite{gao12}. 

\begin{figure}[htb]
\centering
\includegraphics[height=3.05in]{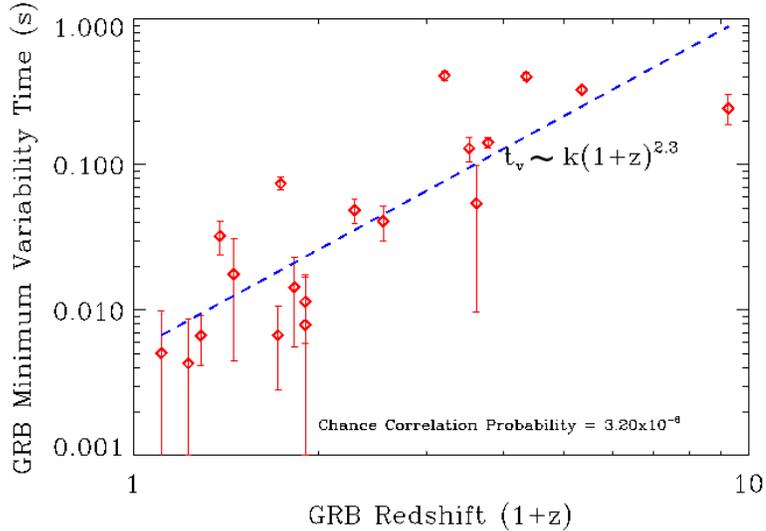}
\caption{: For a small sample of GRBs (both long and short) with known redshifts the 
MVTs are estimated and plotted as a function of (1+z) where z is the source 
redshift. Within statistics MVTs seem to evolve with redshift implying that distant
GRBs have intrinsically longer MVTs.}

\label{fig:rs_mvt}
\end{figure}

If the variability time is a characteristic burst parameter then one expects it be a function of the GRB distance since it will be subjected to cosmological time dilation. However a direct correlation of GRB MVTs with (1+z), where z is the GRB redshift, is in doubt~\cite{koce13}. Figure \ref{fig:rs_mvt} shows a plot of the GRB MVT as a function of (1+z) where z is GRB redshift. The lowest redshift in the present sample is 0.225 for GRB050509B while the highest redshift is 8.1 for GRB090423A. In spite of small sample size there seems to be a good correlation between redshift and MVT with a chance correlation probability of 3.2$\times10^{-6}$.  The variation of the MVT faster than (1+z) shows that there could be a distribution of MVTs at a given redshift.


\section{Discussion}

GRB variability has been used historically to understand the well known compactness problem~\cite{pira99}. Using variability time scales of the GRBs as one of the inputs, minimum bulk Lorentz factors of GRB emission shells have been estimated~\cite{acke10,hasc12, zhao11}. However there was no standard definition nor a standard method of measuring this until recently. The rise time of the narrowest pulse seen in the GRB light curve is logically best measure of the minimum variability time scale since it can be easily be related to values derived using model independent statistical methods like the discrete wavelet transforms~\cite{mach12} or non-parametric analysis using Baysian block techniques~\cite{scar13}.

The new statistical method that has been developed to estimate the minimum variability time scale of a GRB is simple and the feasibility is primarily due to the availability of very high time resolution light curve using the GBM Time Tagged Event (TTE) data. The minimum variability time estimated by this method agrees well with that estimated by discrete wavelet transforms technique as well as the minimum rise time of the fitted lognormal pulses to deconvolve the light curve thus offering a more meaningful interpretation of the burst variability. The minimum variability time can also be used as the optimum time resolution of the light curve before decomposing it with simpler lognormal pulse shapes. In addition, this time scale may be used as a measure of burst variability (V). In other words GRB variability is inversely proportional to the minimum variability time scale, $\tau$. i.e. $V~ \propto ~{{1}\over{\tau}}$

In the past, it has been found that isotropic equivalent peak luminosities, L, of gamma-ray bursts positively correlate with a rigorously constructed measure of the variability of their light curves~i.e. $L ~\propto ~V^{(0.85 \pm 0.02)}$~\cite{reic01,guid05}. We plan to verify this relationship in the future with an enhanced sample size  and using the minimum variability time of each burst as a measure of the its variability.



\begin{thebibliography}{99}


\bibitem{bhat11}
P. N.  Bhat and S. Guiriec, Bull. Astr. Soc. India, {\bf 39}, 471 (2011).
\bibitem{sari97a}
R. Sari and T. Piran, Astrophys. J.. {\bf 485}, 270 (1997).
\bibitem{dermer99}
C. Dermer  and K. E. Mitman, Astrophys. J., {\bf 513}, L5, (1999).
\bibitem{zhang11}
B. Zhang  and  H. Yan, Astrophys. J., {\bf 726}, 90, (2011).
\bibitem{bhat12}
P. N.  Bhat  {\it et al.}, Astrophys. J.. {\bf  744}, 141 (2012).
\bibitem{mach12}
G.  MacLachlan {\it et al.}, Mon. Not. R. Astron. Soc., {\bf 425}, L32 (2012).
\bibitem{gao12}
H. Gao,  {\it et al.}, Astrophys. J.. {\bf  784}, 134 (2012).
\bibitem{sari97b}
R. Sari and T. Piran, Mon. Not. R. Astron. Soc., {\bf 287}, 110 (1997).
\bibitem{koce13}
D. Kocevski  and  V. Petrosian,  Astrophys. J., {\bf 765}, 116, (2013).
\bibitem{pira99}
T. Piran, Phys. Rep., {\bf 314}, 575 (1999).
\bibitem{scar13}
J. D. Scargle  {\it et al.}, Astrophys. J.. {\bf 764}, 167 (2013).
\bibitem{acke10}
M. Ackermann,  {\it et al.}, Astrophys. J.. {\bf 716}, 1178 (2010).
\bibitem{li06}
L. Li and B.  Paczy\'{n}ski, Mon. Not. R. Astron. Soc., {\bf 366}, 219 (2006).
\bibitem{hasc12}
R. Hasco\"{e}t,  {\it et al.}, Mon. Not. R. Astron. Soc., {\bf 421}, 525 (2012).
\bibitem{zhao11}
X. Zhao  {\it et al.}, Astrophys. J.. {\bf 726}, 89 (2011).
\bibitem{reic01}
D. E. Reichart,  {\it et al.}, Astrophys. J.. {\bf 552}, 57 (2001).
\bibitem{guid05}
C. Guidorzi Mon. Not. R. Astron. Soc., {\bf 364}, 163 (2005).

\end{thebibliography}
\end{document}